\documentclass[twocolumn,english,showpacs,preprintnumbers,prb]{revtex4-1}
\usepackage[latin9]{inputenc}
\usepackage{color}
\usepackage{amsmath}
\usepackage{amssymb}
\usepackage{graphicx}
\usepackage{esint}

\makeatletter

\@ifundefined{textcolor}{}{%
 \definecolor{BLACK}{gray}{0}
 \definecolor{WHITE}{gray}{1}
 \definecolor{RED}{rgb}{1,0,0}
 \definecolor{GREEN}{rgb}{0,1,0}
 \definecolor{BLUE}{rgb}{0,0,1}
 \definecolor{CYAN}{cmyk}{1,0,0,0}
 \definecolor{MAGENTA}{cmyk}{0,1,0,0}
 \definecolor{YELLOW}{cmyk}{0,0,1,0}
}

\usepackage{ulem}

\usepackage{aecompl}

\usepackage{epsfig}\usepackage{dcolumn}\usepackage{bm}

\usepackage{babel}

\makeatother

\usepackage{babel}
\begin{document}

\title{Quantum phase transition triggering magnetic BICs in graphene}

\author{L. H. Guessi$^{1}$, Y. Marques$^{2}$, R. S. Machado$^{2}$, K.
Kristinsson$^{3}$,\\
 L. S. Ricco$^{2}$, I. A. Shelykh$^{3,4,5}$, M. S. Figueira$^{6}$,
M. de Souza$^{1,}$}

\altaffiliation{Current address: Institute of Semiconductor and Solid State Physics, Johannes Kepler University Linz, Austria.}

\author{A. C. Seridonio$^{1,2}$}

\affiliation{$^{1}$IGCE, Unesp - Univ Estadual Paulista, Departamento de F\'{i}sica,
13506-900, Rio Claro, SP, Brazil\\
 $^{2}$Departamento de F\'{i}sica e Qu\'{i}mica, Unesp - Univ Estadual
Paulista, 15385-000, Ilha Solteira, SP, Brazil\\
 $^{3}$Division of Physics and Applied Physics, Nanyang Technological
University 637371, Singapore\\
 $^{4}$Science Institute, University of Iceland, Dunhagi-3, IS-107,
Reykjavik, Iceland\\
 $^{5}$ITMO University, St. Petersburg 197101, Russia\\
 $^{6}$Instituto de F\'{i}sica, Universidade Federal Fluminense,
24210-340, Niterói, RJ, Brazil}
\begin{abstract}
Graphene hosting a pair of collinear adatoms in the phantom atom configuration
has density of states vanishing in the vicinity of the Dirac point
which can be described in terms of the pseudogap scaling as cube of
the energy, $\Delta\propto|\varepsilon|^{3}$ which leads to the appearance
of spin-degenerate bound states in the continuum (BICs) {[}Phys. Rev.
B \textbf{92}, 045409 (2015){]}. In the case when adatoms are locally
coupled to a single carbon atom the pseudogap scales linearly with
energy, which prevents the formation of BICs. Here, we explore the
effects of non-local coupling characterized by the Fano factor of
interference $q_{0},$ tunable by changing the slope of the Dirac
cones in the graphene band-structure. We demonstrate that three distinct
regimes can be identified: i) for $q_{0}<q_{c1}$ (critical point)
a mixed pseudogap $\Delta\propto|\varepsilon|,|\varepsilon|^{2}$
appears yielding a phase with spin-degenerate BICs; ii) near $q_{0}=q_{c1}$
when $\Delta\propto|\varepsilon|^{2}$ the system undergoes a quantum
phase transition (QPT) in which the new phase is characterized by
magnetic BICs and iii) at a second critical value $q_{0}>q_{c2}$
the cubic scaling of the pseudogap with energy $\Delta\propto|\varepsilon|^{3}$
characteristic to the phantom atom configuration is restored and the
phase with non-magnetic BICs is recovered. The phase with magnetic
BICs can be described in terms of an effective intrinsic exchange
field of ferromagnetic nature between the adatoms mediated by graphene
monolayer. We thus propose a new type of QPT resulting from the competition
between two ground states, respectively characterized by spin-degenerate
and magnetic BICs.
\end{abstract}

\pacs{72.80.Vp, 07.79.Cz, 72.10.Fk}

\maketitle

\section{Introduction}

Graphene-based systems are promising candidates for the detection
of the so-called bound states in the continuum (BICs) \cite{LH,Gong}.
BICs were first theoretically predicted by von Neumann and Wigner
in 1929 \cite{Neuman-1} as quantum states with localized square-integrable
wave functions, but having energies within the continuum of delocalized
states. The electrons within BICs do not decay into the system continuum,
thus these states should be invisible in transport experiments.

The subject experienced revival after the work of Stillinger and Herrick
in 1975 \cite{SH}. Since then, BICs were predicted to appear in a
variety of electronic, optical and photonic systems \cite{LH,Boretz,Crespi}.
In these systems, effects of Fano interference \cite{Fano1} were
proposed as the underlying mechanism for the emergence of BICs and
their possible experimental observation. In particular, we recently
proposed that BICs can be observed in the system of graphene with
two collinear adatoms in the phantom atom configuration \cite{LH}.

\begin{figure}[!]
\includegraphics[width=0.48\textwidth,height=0.21\textheight]{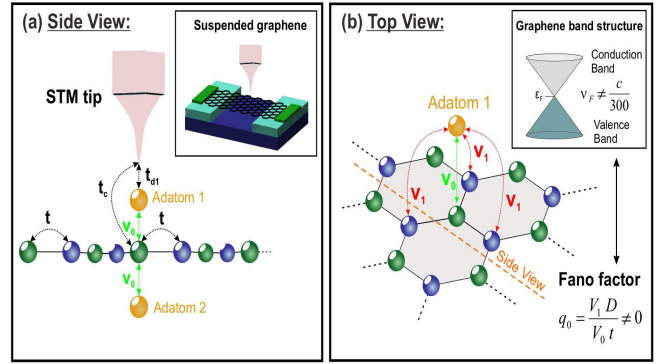}
\protect\protect\protect\protect\protect\protect\protect\protect\protect\protect\protect\protect\caption{\label{fig:PPic1} (Color online) (a) Side view: two adatoms labeled
by $1$ (upper) and $2$ (lower) placed collinear to a carbon atom
beneath an STM tip in suspended graphene (inset) (b) Top view: the
adatoms are coupled to the carbon atom beneath them and its nearest
neighbors. The relative strength of these couplings define the Fano
factor of interference $q_{0}$ playing the role of the natural control
parameter of the system. It can be tuned by varying the slope of the
Dirac cones in the graphene band-structure (inset).}
\end{figure}

In this work, we show that the setup outlined in Fig.\ref{fig:PPic1}
for suspended graphene can undergo a quantum phase transition (QPT)
into the state with magnetic BICs if non-local graphene-adatom couplings
are taken into account. The phenomenon is a consequence
of the particular scaling of the local density of states (LDOS) $\text{\ensuremath{\mathcal{D}}}_{0}$
on energy $\varepsilon$ in the vicinity of the Dirac point. The latter
is proportional to the quantity known as \textit{pseudogap}
$\Delta$, related to the intensity of the scattering near the Fermi
energy \cite{Gregorio,Ingersent}. Formation of the magnetic BICs
becomes possible only if $\Delta\propto|\varepsilon|^{2}$ similar
to the transition reported in Ref.\,{[}\onlinecite{Gregorio}{]}
for a pair of quantum dots coupled to metallic leads.

The QPT reported here is driven by a Fano factor of interference $q_{0}$
which can be thus considered as the natural control parameter of the
system. It can be tuned by changing the slope of the Dirac cones in
the graphene band-structure (see Fig.\ref{fig:PPic1}(b)). The magnetic
BICs appear within the region inside the critical boundaries $q_{c1}<q_{0}<q_{c2}$,
where the dominant scaling law for the pseudogap is quadratic ($\Delta\propto|\varepsilon|^{2}$).
Outside this region, the mixed scaling $\Delta\propto|\varepsilon|,|\varepsilon|^{2}$
for $q_{0}<q_{c1}$ or cubic scaling $\Delta\propto|\varepsilon|^{3}$
for $q_{0}>q_{c2},$ leads to the formation of spin-degenerate BICs.
The transition towards the magnetic BIC state is triggered due to
the onset of the effective intrinsic ferromagnetic exchange field
$\mathcal{J}^{\text{{exch}}}$ between the adatoms mediated by the
graphene monolayer.

\section{The model}

To give a theoretical description of the setup plotted in Fig.\ref{fig:PPic1},
we use the model based on the Anderson Hamiltonian \cite{Anderson1,Anderson2}:
\begin{align}
\mathcal{H}_{2D} & =\sum_{s\sigma}\int dk(\hbar v_{F}k)c_{sk\sigma}^{\dagger}c_{sk\sigma}+\sum_{j\sigma}\varepsilon_{jd}d_{j\sigma}^{\dagger}d_{j\sigma}\nonumber \\
 & +\mathcal{U}\sum_{j}n_{d_{j}\uparrow}n_{d_{j}\downarrow}+\sum_{js\sigma}\int dk\mathcal{V}_{k}(c_{sk\sigma}^{\dagger}d_{j\sigma}+\text{H.c.}),\label{eq:TIAM}
\end{align}
with $v_{F}$ being Fermi velocity. The graphene monolayer is described
by operators $c_{sk\sigma}^{\dagger}$ ($c_{sk\sigma}$) for creation
(annihilation) of electrons in quantum states labeled by the wave
number $k$, spin $\sigma$ and valley index $s=1,2$. For the adatoms,
$d_{j\sigma}^{\dagger}$ ($d_{j\sigma}$) creates (annihilates) an
electron with spin $\sigma$ with energy $\varepsilon_{jd}$, where
$j=1,2$ correspond to the upper and lower adatoms, respectively.
The third term in Eq.(\ref{eq:TIAM}) accounts for the on-site Coulomb
interaction $\mathcal{U}$, with $n_{d_{j}{\sigma}}=d_{j\sigma}^{\dagger}d_{j\sigma}$.
Finally, the last term mixes the graphene and the levels $\varepsilon_{jd}$,
wherein $\text{H.c.}$ gives the Hermitian conjugate of the first
part. This mixing is characterized by the coupling $\mathcal{V}_{k}=\frac{1}{2\pi}\sqrt{\frac{\pi\Omega_{0}}{\mathcal{N}}}\sqrt{\left|k\right|}v_{0}(1-q_{0}\frac{\hbar v_{F}k}{D}),$
where $\mathcal{N}$ is the number of conduction states, $\Omega_{0}$
denotes the unit cell area, and
\begin{equation}
q_{0}=\frac{v_{1}D}{v_{0}t}
\end{equation}
is the Fano factor of interference defined according to the results
of Ref.\,{[}\onlinecite{Tshaped}{]}. The parameter $t$ stands
for the coupling strength between carbon atoms, while $v_{0}$ and
$v_{1}$ represent the host-adatom hybridizations outlined in Fig.\,\ref{fig:PPic1}
and $D=7\text{{eV}}$ denotes the band-edge for $v_{F}\sim c/300$.
The Fano factor $q_{0}$ can be tuned assisted by a variation of $v_{F},$
which enters into $t=\frac{2\hbar}{3a}v_{F}$ \cite{CNeto1} and $\frac{v_{1}D}{v_{0}}$.
The experimental tuning of $v_{F}$ can be achieved, for instance,
by means of modifying the carrier concentration in suspended graphene
\cite{TFV2} {[}inset of Fig.\,\ref{fig:PPic1}(a){]}.

The situation $q_{0}=0$ corresponds to the scenario in which collinear
adatoms are locally side-coupled to a single carbon atom (local coupling
regime). Otherwise, $q_{0}\neq0$ denotes the hybridization of the
adatoms with the three second neighbors of carbons as depicted in
Fig.\,\ref{fig:PPic1} (non-local coupling).

To analyze the transport properties of the geometry
we consider and look for the existence of the BICs, we should focus
on the local density of states of the host (LDOS) and those corresponding for the adatoms (DOS).
The former defines the conductance of the device at zero temperature $T=0$\cite{LH}:
\begin{equation}
G\sim\frac{e^{2}}{h}\Gamma_{\text{{tip}}}\text{{LDOS}},\label{eq:Conductance}
\end{equation}
with $\Gamma_{\text{{tip}}}=\pi t_{c}^{2}\rho_{\text{{tip}}},$ $\rho_{\text{{tip}}}$
as the STM tip density of states.

To obtain the value of LDOS probed by the STM tip
of Fig.\,\ref{fig:PPic1}, we should consider the tunneling Hamiltonian
\begin{equation}
\mathcal{H}_{\text{tun}}=t_{c}\sum_{\sigma}\psi_{\text{tip},\sigma}^{\dagger}\Psi_{\sigma}+\text{H.c.},
\end{equation}
where $\psi_{\text{tip},\sigma}$ and $\Psi_{\sigma}$ are respectively
fermionic operators for the edge site of the STM tip and
\begin{align}
\Psi_{\sigma} & =\frac{1}{2\pi}\sqrt{\frac{\pi\Omega_{0}}{\mathcal{N}}}\sum_{s}\int\sqrt{\left|k\right|}(1-q_{0}\frac{\hbar v_{F}k}{D})dkc_{sk\sigma}+\frac{t_{d_{1}}}{t_{c}}d_{1\sigma}\label{eq:FOperator}
\end{align}
is the field operator accounting for the quantum state of the graphene
site placed right beneath the tip with hopping terms ($t_{d_{1}}$
and $t_{c}$), cf. Fig.\,\ref{fig:PPic1}. LDOS then can be computed
as
\begin{equation}
\text{{LDOS}}=-\frac{1}{\pi}\sum_{\sigma}{\tt Im}[\tilde{\mathcal{G}}_{\sigma}(\varepsilon^{+})]=2\text{\ensuremath{\mathcal{D}}}_{0}+\sum_{\sigma jl}\Delta\text{{LDOS}}_{jl\sigma},\label{eq:FM_LDOS}
\end{equation}
where $\tilde{\mathcal{G}}_{\sigma}(\varepsilon^{+})$ is the time
Fourier transform of the Green's function
\begin{equation}
\mathcal{G}_{\sigma}=-\frac{i}{\hbar}\theta\left(\tau\right){\tt Tr}\{\varrho_{\text{2D}}[\Psi_{\sigma}(\tau),\Psi_{\sigma}^{\dagger}(0)]_{+}\}
\end{equation}
and
\begin{equation}
\text{\ensuremath{\mathcal{D}}}_{0}=\frac{\left|\varepsilon\right|}{D^{2}}(1-q_{0}\frac{\varepsilon}{D})^{2}\label{D0}
\end{equation}
is the graphene DOS, $\Delta\text{{LDOS}}_{jl\sigma}$ stands for the part induced
by the adatoms (see detailed derivation for it in the Appendix).

It is worth mentioning that $\Delta\text{LDOS}_{jl\sigma}$
for $j\neq l$ represents electronic waves of a given spin $\sigma$
that travel forth and back between the upper and lower adatoms showed
in Fig.\,\ref{fig:PPic1}(a), which for a specific energy $\varepsilon,$
become phase shifted by $\pi$ (Fano dip) with respect to the waves
scattered by the adatoms, which are described by $\Delta\text{LDOS}_{jj\sigma}$.
As discussed in Ref.{[}\onlinecite{LH}{]}, such scattering process then provides
a mechanism of the emergence of BICs. This effect can be captured
in the detailed derivation of LDOS appearing in the Appendix.

According to the Appendix, the evaluation of $\Delta\text{LDOS}_{jl\sigma}$ depends on the Green's functions $\tilde{\mathcal{G}}_{d_{j\sigma}d_{l\sigma}}$  ($j=1,2$ and $l=1,2$ ) for the adatoms. Additionally, to perceive the BICs emergence in our system, we should know the density of states $\text{DOS}_{jj}^{\sigma}$ of
these adatoms, which are determined as follows:
\begin{equation}
\text{DOS}_{jj}^{\sigma}=-\frac{1}{\pi}{\tt Im}(\tilde{\mathcal{G}}_{d_{j\sigma}d_{j\sigma}}).\label{eq:DOSjj}
\end{equation}
Thus both $\Delta\text{LDOS}_{jl\sigma}$ and $\text{DOS}_{jj}^{\sigma}$ can be found by employing the Hubbard I approach \cite{Hubbard} at $T=0$,
since the determined Hubbard bands match with those obtained via the
Numerical Renormalization Group, in particular, for graphene with
a single adatom \cite{Balseiro}. As a result, we can safely extrapolate
the Hubbard I method to our graphene system. We start employing the
equation-of-motion method to a single particle retarded Green's function
of an adatom in time domain $\mathcal{G}_{d_{l\sigma}d_{j\sigma}}=-\frac{i}{\hbar}\theta\left(\tau\right){\tt Tr}\{\varrho_{\text{2D}}[d_{l\sigma}\left(\tau\right),d_{j\sigma}^{\dagger}\left(0\right)]_{+}\},$
where $\theta\left(\tau\right)$ is the Heaviside function, $\varrho_{\text{2D}}$
is the density matrix of the system described by the Hamiltonian of
Eq.(\ref{eq:TIAM}) and $[\cdots,\cdots]_{+}$ is the anticommutator
between operators taken in the Heisenberg picture. Performing elementary
algebra one obtains in the energy domain:
\begin{align}
(\varepsilon^{+}-\varepsilon_{ld})\tilde{\mathcal{G}}_{d_{l\sigma}d_{j\sigma}} & =\delta_{lj}+\Sigma\sum_{\tilde{l}}\tilde{\mathcal{G}}_{d_{\tilde{l}\sigma}d_{j\sigma}}\nonumber \\
 & +\mathcal{U}\tilde{\mathcal{G}}_{d_{l\sigma}n_{d_{l}\bar{\sigma}},d_{j\sigma}},\label{eq:s1}
\end{align}
where $\varepsilon^{+}=\varepsilon+i0^{+}$ and
\begin{align}
\Sigma & =\sum_{s}\int dk\frac{\mathcal{V}_{k}\mathcal{V}_{k}}{\varepsilon^{+}-\hbar v_{F}k}=-\frac{v_{0}^{2}}{D^{2}}\varepsilon(1-q_{0}\frac{\varepsilon}{D})^{2}\text{ln}\left|\frac{D^{2}-\varepsilon^{2}}{\varepsilon^{2}}\right|\nonumber \\
 & +\frac{v_{0}^{2}}{D}q_{0}(2-q_{0}\frac{\varepsilon}{D})-i\Delta\label{eq:SE}
\end{align}
is the self-energy. Its imaginary part $\Delta$ is
proportional to the scattering rate of the quasiparticles and is known
as pseudogap. The latter is proportional to the local density of states
of the host $\text{\ensuremath{\mathcal{D}}}_{0}$ given by Eq.(\ref{D0}),
i.e., $\Delta=\pi v_{0}^{2}\text{\ensuremath{\mathcal{D}}}_{0}$ \cite{Gregorio,Ingersent}.
Thus
\begin{equation}
\Delta=\frac{\pi v_{0}^{2}}{D^{2}}\left|\varepsilon\right|(1-q_{0}\frac{\varepsilon}{D})^{2},\label{eq:broad}
\end{equation}
which depending on the value of the Fano parameter,
the main contribution to the pseudogap can be linear, cubic or quadratic.
As we will see in the discussion section, the latter situation is
of particular interest, since magnetic BICs are formed in this case.

In Eq.\,(\ref{eq:s1}) $\tilde{\mathcal{G}}_{d_{l\sigma}n_{d_{l}\bar{\sigma}},d_{j\sigma}}$
is a two particle Green's function composed by four fermionic operators,
obtained from the time Fourier transform of $\mathcal{G}_{d_{l\sigma}n_{d_{l}\bar{\sigma}},d_{j\sigma}}=-\frac{i}{\hbar}\theta(\tau){\tt Tr}\{\varrho_{\text{2D}}[d_{l\sigma}\left(\tau\right)n_{d_{l}\bar{\sigma}}\left(\tau\right),d_{j\sigma}^{\dagger}\left(0\right)]_{+}\},$
with $n_{d_{l}\bar{\sigma}}=d_{l\bar{\sigma}}^{\dagger}d_{l\bar{\sigma}}$
and spin $\bar{\sigma}$ (opposite to $\sigma$). Thus we first calculate
the time derivative of $\mathcal{G}_{d_{l\sigma}n_{d_{l}\bar{\sigma}},d_{j\sigma}}$
and then its time Fourier transform, which leads to
\begin{align}
(\varepsilon^{+}-\varepsilon_{ld}-\mathcal{U})\tilde{\mathcal{G}}_{d_{l\sigma}n_{d_{l}\bar{\sigma}},d_{j\sigma}} & =\delta_{lj}<n_{d_{l}\bar{\sigma}}>\nonumber \\
+\sum_{s}\int dk\mathcal{V}_{k} & (\tilde{\mathcal{G}}_{c_{sk\sigma}d_{l\bar{\sigma}}^{\dagger}d_{l\bar{\sigma}},d_{j\sigma}}\nonumber \\
-\tilde{\mathcal{G}}_{c_{sk\bar{\sigma}}^{\dagger}d_{l\bar{\sigma}}d_{l\sigma},d_{j\sigma}} & +\tilde{\mathcal{G}}_{d_{l\bar{\sigma}}^{\dagger}c_{sk\bar{\sigma}}d_{l\sigma},d_{j\sigma}}),\label{eq:H_GF_2}
\end{align}
expressed in terms of new Green's functions of the same order of $\tilde{\mathcal{G}}_{d_{l\sigma}n_{d_{l}\bar{\sigma}},d_{j\sigma}}$
and the occupation number $<n_{d_{l}\bar{\sigma}}>$ determined by
\begin{equation}
<n_{d_{l}\bar{\sigma}}>=-\frac{1}{\pi}\int_{-D}^{\varepsilon_{F}=0}{\tt Im}(\tilde{\mathcal{G}}_{d_{l\bar{\sigma}}d_{l\bar{\sigma}}})d\varepsilon.\label{eq:nb}
\end{equation}

We highlight that for the quadratic pseudogap, the
self-consistent evaluation of the Eq.\,(\ref{eq:nb}) reveals a range
of magnetic solutions with $<n_{d_{l}\uparrow}>\neq<n_{d_{l}\downarrow}>$
for the values of $q_{0}$, lying in the range between two critical
points $q_{c1}$ and $q_{c2}$. Outside the magnetic region, i.e.,
for different scalings of the pseudogap ($\Delta\propto|\varepsilon|,|\varepsilon|^{2}$
for $q_{0}<q_{c1}$ and $\Delta\propto|\varepsilon|^{3}$ for $q_{0}>q_{c2}$),
Eq.\,(\ref{eq:nb}) has non-magnetic solutions with $<n_{d_{l}\uparrow}>=<n_{d_{l}\downarrow}>$
only. This point will be addressed in detail in Sec.III of the paper
(see in particular Fig.\ref{fig:PPic2}).

Furthermore, by employing the Hubbard I approximation, we decouple
the Green's functions in the right-hand side of Eq.\,(\ref{eq:H_GF_2})
as performed in Ref.\,{[}\onlinecite{LH}{]}. This procedure enables
us to solve the system of Green's functions within Eq.\,(\ref{eq:s1}),
leading to $\tilde{\mathcal{G}}_{d_{j\sigma}d_{j\sigma}}=\frac{\lambda_{j}^{\bar{\sigma}}}{\varepsilon-\varepsilon_{jd}-{{\tilde{\Sigma}}^{\bar{\sigma}}}_{j\bar{j}}},$
where $\lambda_{j}^{\bar{\sigma}}=(1+\frac{\mathcal{U}<n_{d_{j}\bar{\sigma}}>}{\varepsilon-\varepsilon_{jd}-\mathcal{U}-\Sigma})$,
and:
\begin{equation}
{{\tilde{\Sigma}}^{\bar{\sigma}}}_{j\bar{j}}=\Sigma+\lambda_{j}^{\bar{\sigma}}\lambda_{\bar{j}}^{\bar{\sigma}}\frac{\Sigma^{2}}{\varepsilon-\varepsilon_{\bar{j}d}-\Sigma}\label{eq:TSE}
\end{equation}
is the total self-energy, with $\bar{j}=2,1$ respectively for $j=1,2$
in order to identify distinct adatoms and $\tilde{\mathcal{G}}_{d_{j\sigma}d_{\bar{j}\sigma}}=\frac{\lambda_{j}^{\bar{\sigma}}\Sigma\tilde{\mathcal{G}}_{d_{\bar{j}\sigma}d_{\bar{j}\sigma}}}{\varepsilon-\varepsilon_{jd}-\Sigma}$
are mixed Green's functions.

\section{Results and Discussion}

In the simulations we adopt $T=0$ and the set of parameters \cite{LH}:
$\varepsilon_{jd}=\varepsilon_{d}=-0.07D$, which is feasible in suspended
graphene (inset of Fig.\,\ref{fig:PPic1}(a)) \cite{TFV2} and $\mathcal{U}=v_{0}=-2\varepsilon_{d}.$
Additionally, to avoid that BICs decay into the continuum, we use
$t_{d_{1}}/t_{c}=0$, otherwise it leads to experimental detection
of BICs by means of the so-called quasi-BICs \cite{LH}.

\begin{figure}
\includegraphics[width=0.45\textwidth,height=0.3\textheight]{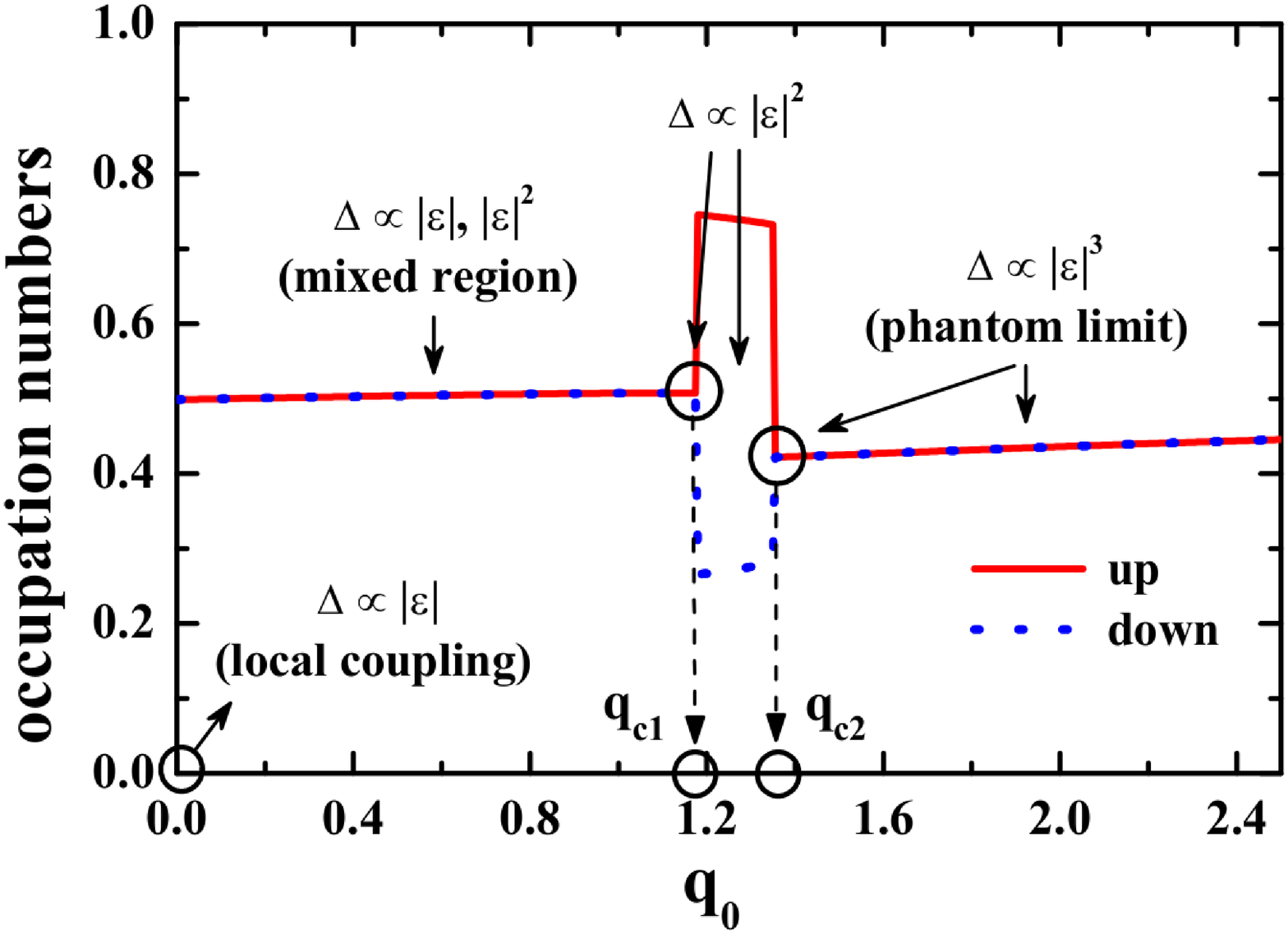}
\protect\protect\protect\protect\protect\protect\protect\protect\protect\protect\protect\protect\caption{\label{fig:PPic2}(Color online) Occupation numbers given by Eq.(\ref{eq:nb})
for spin-up and spin-down states of the adatoms as a function of $q_{0}$. }
\end{figure}

\begin{figure}
\includegraphics[width=0.45\textwidth,height=0.33\textheight]{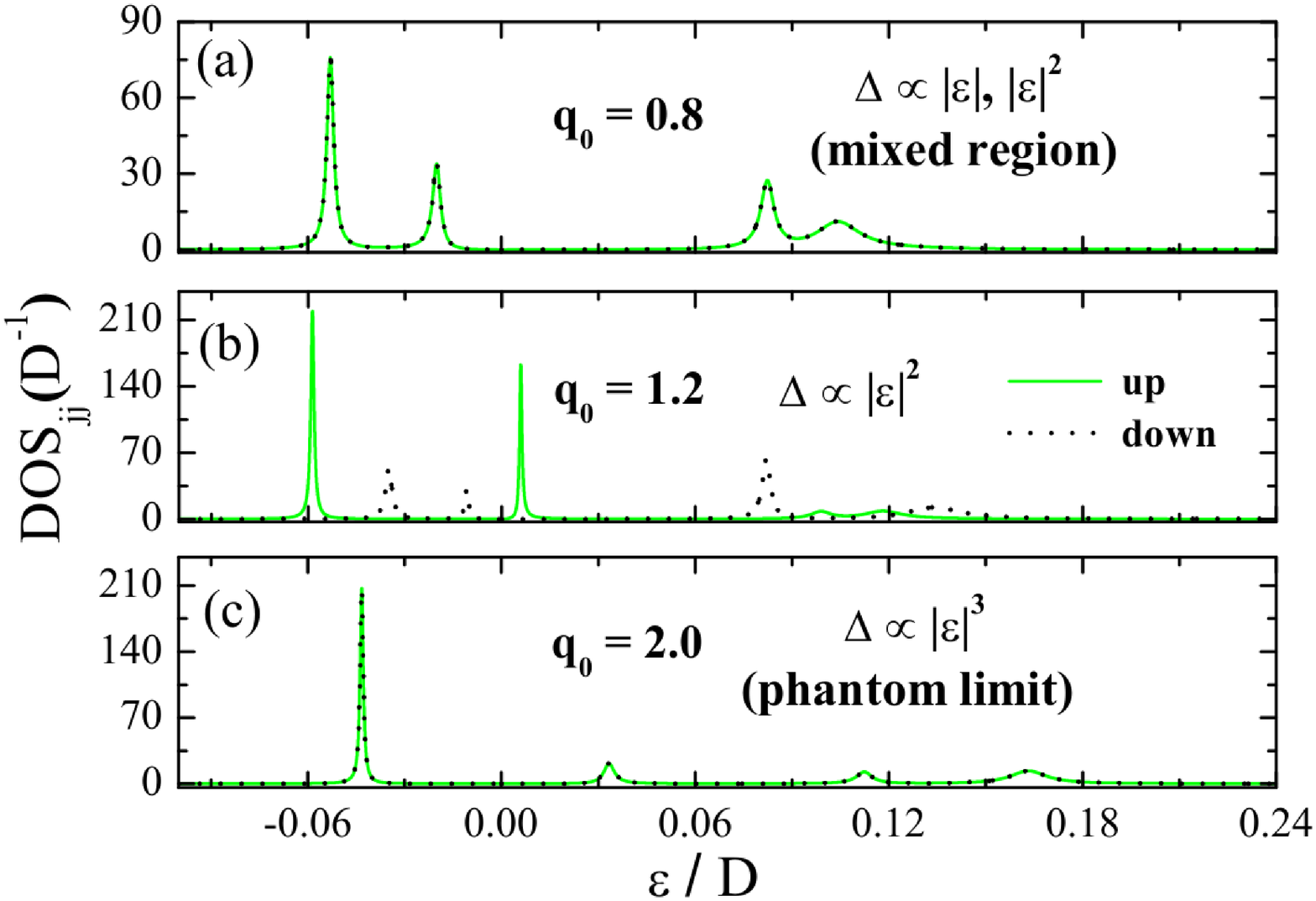}
\protect\protect\protect\protect\protect\protect\protect\protect\protect\protect\protect\protect\caption{\label{fig:PPic3}(Color online) (a) DOS for the case $q_{0}<q_{c1}$.
Well resolved spin degenerate peaks are clearly visible. (b) DOS for
the case $q_{c1}<q_{0}<q_{c2}$ with break of spin degeneracy. (c)
DOS for the case $q_{0}>q_{c2}$ when spin degeneracy is recovered.}
\end{figure}

In Fig.\,\ref{fig:PPic2} three distinct regions in the occupation
numbers of Eq.\,(\ref{eq:nb}) for $j=1,2$ appear identified by
their corresponding pseudogaps $\Delta$ {[}Eq.\,(\ref{eq:broad}){]}:
the non-magnetic regions corresponding to small or big Fano factors
appear to be divided by a magnetic central domain delimited by the
critical values $q_{c1}$ and $q_{c2}.$ At critical values, abrupt
jumps in the occupation numbers point out the existence of a QPT connected
with the spin degree of freedom. Panel (a) of Fig.\,\ref{fig:PPic3}
presents the DOS corresponding to the regime $q_{0}=0.8<q_{c1}$ where
one can clearly see resolved and spin-degenerate peaks in Eq.\,(\ref{eq:DOSjj})
for the $\text{DOS}_{jj}^{\sigma}.$ In Fig.\,\ref{fig:PPic3}(b)
spin-polarized peaks emerge when the Fano factor is placed within
the boundaries $q_{c1}<q_{0}=1.2<q_{c2},$ while in panel (c) the
case of $q_{0}=2.0$ corresponds to the limit of the phantom atom
considered in detail in Ref.{[}\onlinecite{LH}{]} for which spin
degeneracy is recovered.

\begin{figure}
\includegraphics[width=0.45\textwidth,height=0.33\textheight]{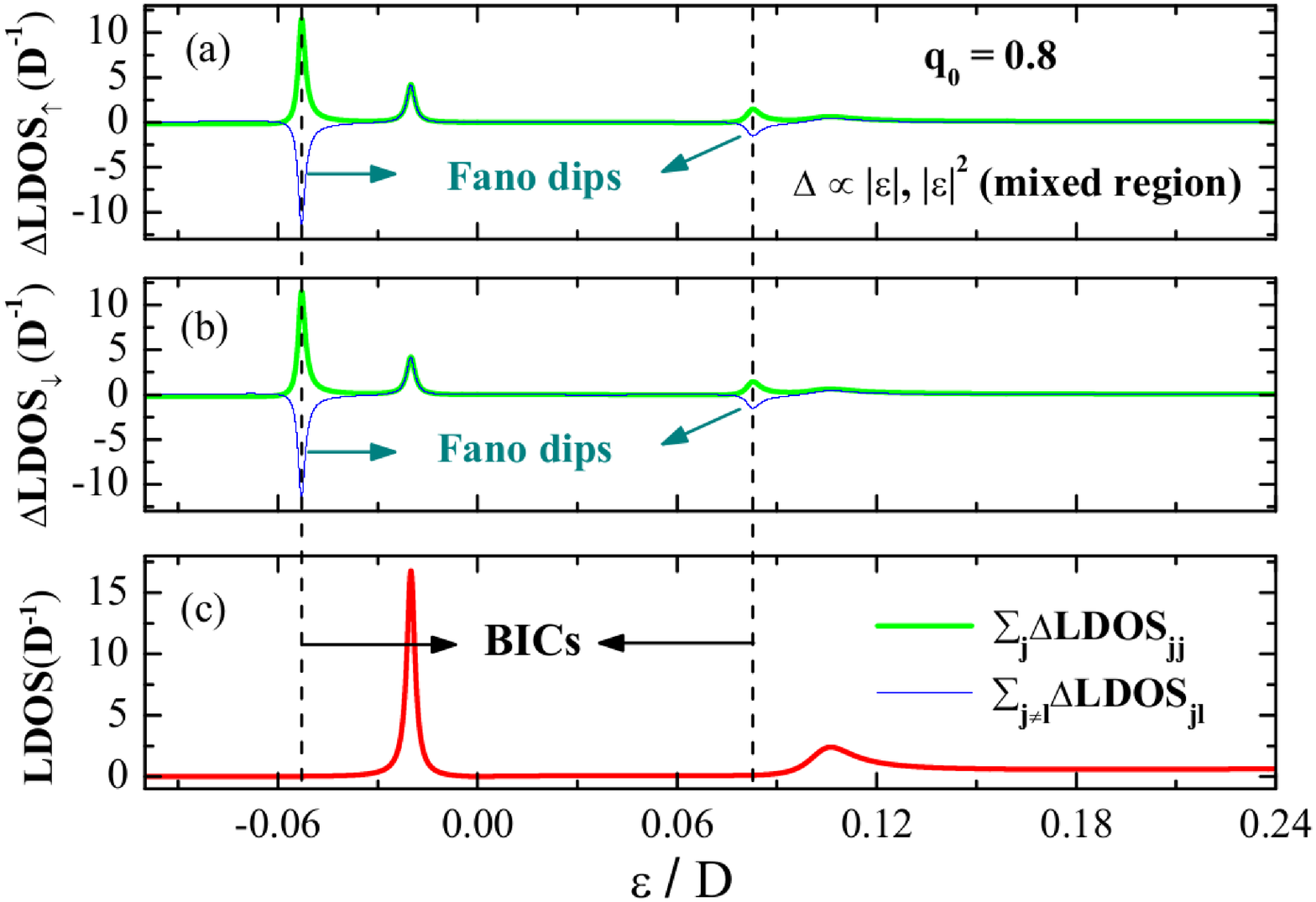}
\protect\protect\protect\protect\protect\protect\protect\protect\protect\protect\protect\protect\caption{\label{fig:PPic4}(Color online) Host local density of states corresponding
to the cases of non-magnetic BICs (panels (a), (b), and (c)). BICs
appear when a peak in $\Sigma_{j}\Delta\text{LDOS}_{jj\sigma}$ is
fully compensated by a Fano dip in $\Sigma_{j\protect\neq l}\Delta\text{LDOS}_{jl\sigma}$.
Positions of BICs are marked by vertical dashed lines. Panels (a)
and (b) correspond to spin resolved $\Delta\text{LDOS}$. Lower panel
(c) corresponds to total LDOS defining the conductance.}
\end{figure}

To demonstrate that the system possesses BICs, we compare the density
of states $\text{DOS}_{jj}^{\sigma}$ for adatoms shown in Fig.\ref{fig:PPic3}
with the host local density of states $\Delta\text{{LDOS}}{}_{jj\sigma}$
depicted at Fig.\ref{fig:PPic4}. As one can see, both quantities
reveal pronounced peaks (resonant states) placed at the same positions.
Particularly in panels (a) and (b) of Fig.\,\ref{fig:PPic4} with
$q_{0}=0.8$, we observe as aftermath of Eq.(\ref{eq:LDOSp1}), degenerate
spin-up and down components for the Fano dips of $\Delta\text{{LDOS}}{}_{jl\sigma}$
$(l\neq j)$ interfering destructively with the peaks found in $\Delta\text{{LDOS}}{}_{jj\sigma}.$
As this interference is completely perfect, BICs emerge at the positions
marked by vertical lines crossing panels (a), (b) and (c) of this
figure. In panel (c) of the same figure, the total LDOS of Eq.\,(\ref{eq:FM_LDOS})
reveals absence of peaks at those places in which such a destructive
interference occurs within panels (a) and (b). The aforementioned
positions without peaks in Fig.\,\ref{fig:PPic4}(c) thereby give
rise to BICs: the total LDOS that determines the conductance does
not catch the same peaks found in Fig.\,\ref{fig:PPic3}(a) for the
adatoms. Thus the aforementioned invisibility of such resonant states
points out that electrons with opposite spins stay equally trapped
within these adatoms when $q_{0}<q_{c1}$ and the pseudogap scales
as $\Delta\propto|\varepsilon|,|\varepsilon|^{2}.$

\begin{figure}
\includegraphics[width=0.46\textwidth,height=0.33\textheight]{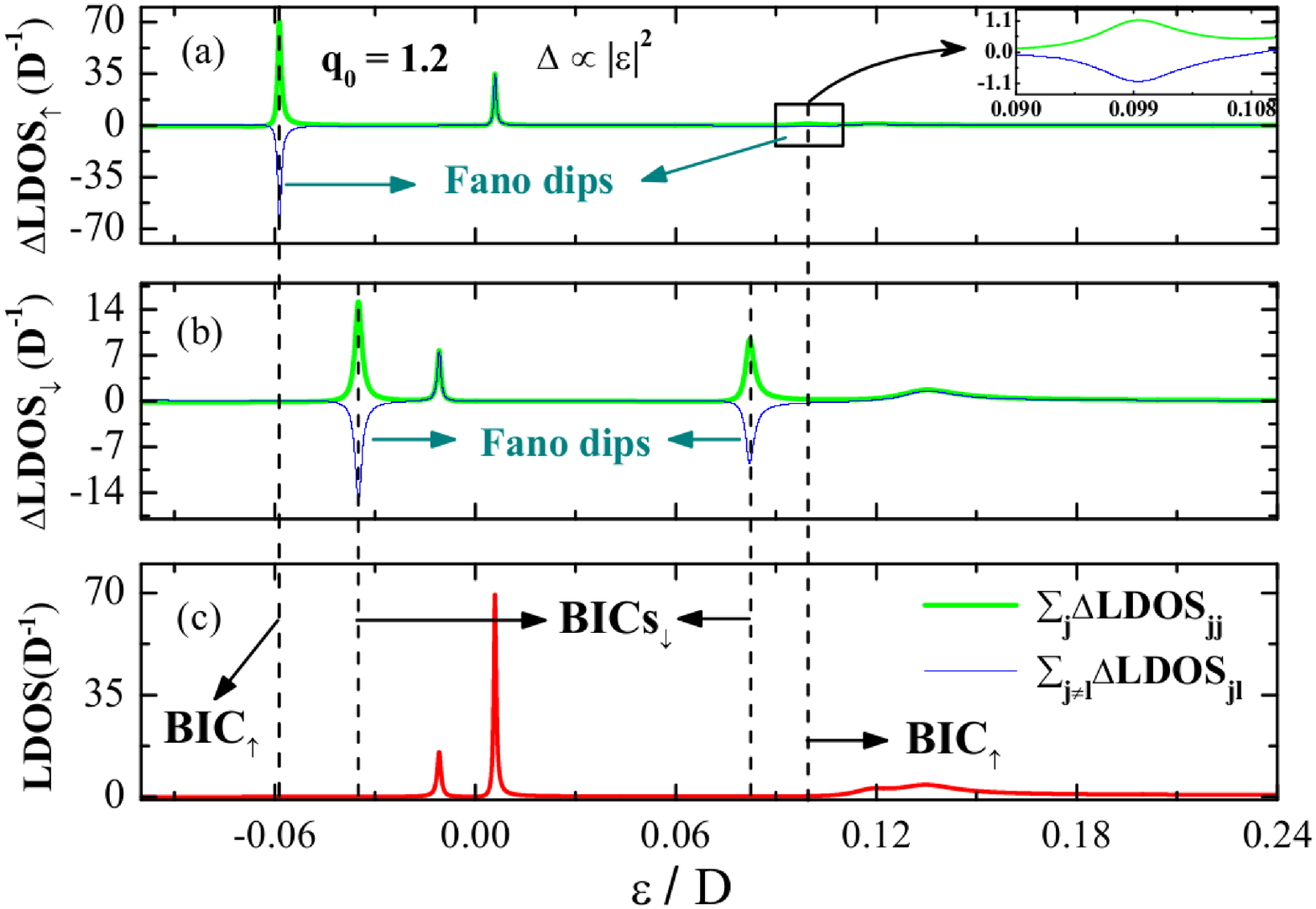}
\protect\protect\protect\protect\protect\protect\protect\protect\protect\protect\protect\protect\caption{\label{fig:PPic5}(Color online) Host local density of states corresponding
to the cases of magnetic BICs (panels (a), (b), and (c)).}
\end{figure}

Panels (a), (b) and (c) of Fig.\,\ref{fig:PPic5} correspond to the
case $q_{c1}<q_{0}=1.2<q_{c2}$ where magnetic solutions become possible,
since the pseudogap is ruled by $\Delta\propto|\varepsilon|^{2}$.
The position of magnetic BICs is denoted by vertical dashed lines.
Consequently, in the domain $q_{c1}<q_{0}<q_{c2},$ the novelty due
to a non-local coupling between graphene and collinear adatoms lies
on the possibility of tuning the spin of the electrons trapped in
the BICs of the adatoms. Such a feature yields an emerging \textit{based
suspended graphene spintronics}, in which a spin-filter of BICs rises
as a feasible application. Outside the critical domain, just spin-degenerate
BICs exist.

Let us now present the physical arguments that elucidate the emergence
of the reported magnetic BICs, which is indeed triggered by a QPT.
Similar QPT appearing due to the quadratic scaling of the pseudogap
$\Delta\propto|\varepsilon|^{2}$ and related breaking of the spin-degeneracy
was discussed in Ref.\,{[}\onlinecite{Gregorio}{]}, where a double
dot system was explored. In regard of this dot setup, we highlight
that the pseudogap $\Delta\propto|\varepsilon|^{2}$ is only revealed
to be present after performing a mapping of the original Hamiltonian
into an effective model, in particular, under restricted constraints.
On the other hand, we demonstrate that graphene emerges as the natural
platform wherein the pseudogap of Eq.(\ref{eq:broad}) includes not
only the regime $|\varepsilon|^{2}$, but also $|\varepsilon|,|\varepsilon|^{2}$
and $|\varepsilon|^{3}$, just due to the non-local adatom-graphene
coupling. These regimes are accessible by means of the tuning of the
Fano factor $q_{0}$, which here is proposed to be practicable by
developing the Fermi velocity engineering \cite{TFV2}. Moreover,
the non-local coupling assumption improves the emulation of the experimental
reality, since the standard case of local coupling regime, which is
widely employed in the literature, is indeed ideal and hides completely
the reported QPT.

\begin{figure}
\includegraphics[width=0.45\textwidth,height=0.31\textheight]{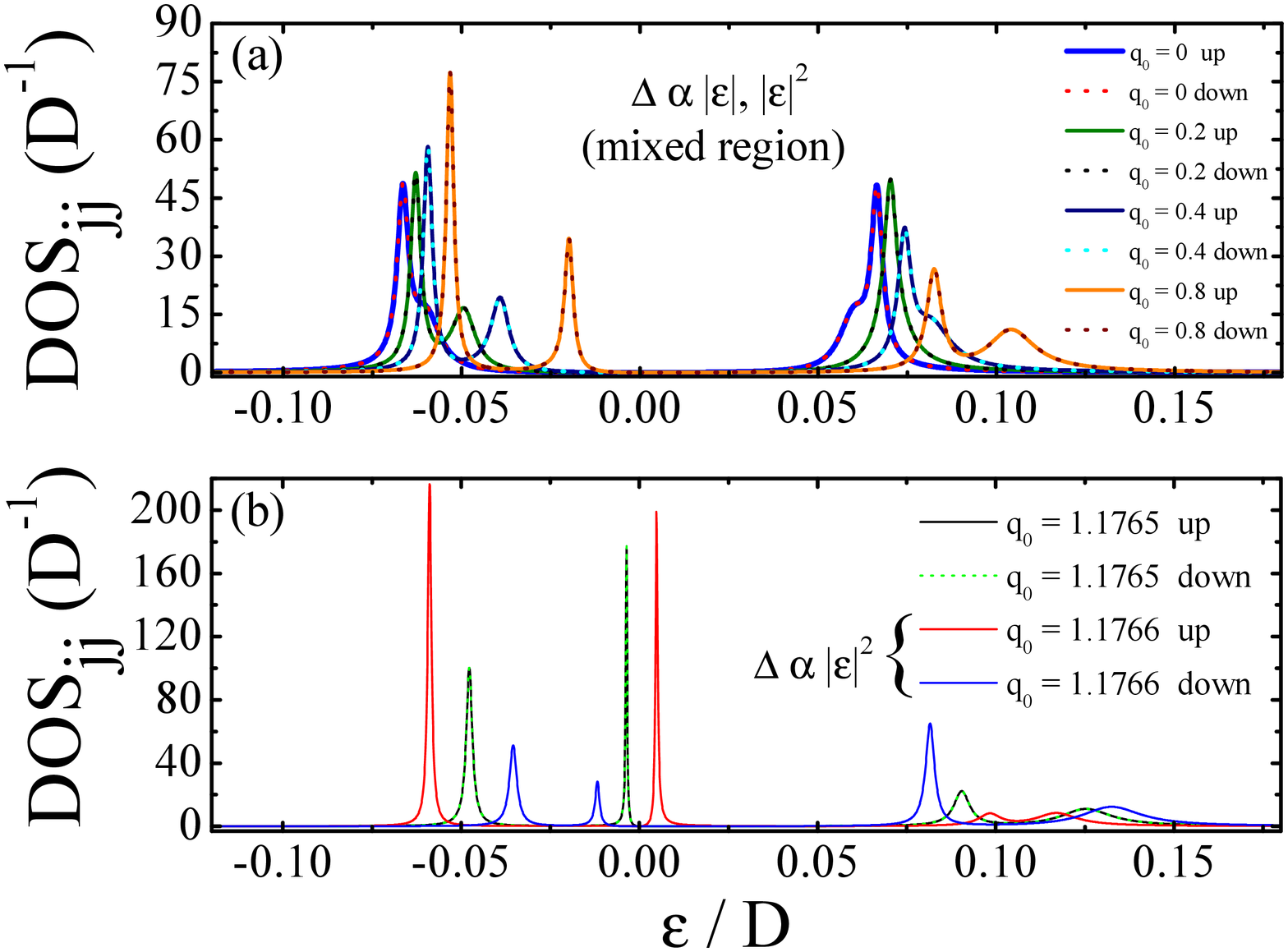}
\protect\protect\protect\protect\protect\protect\protect\protect\protect\protect\protect\protect\caption{\label{fig:PPic6}(Color online) (a) Spin-degenerate crossover from
merged peaks of Eq.(\ref{eq:DOSjj}) for the adatoms DOS towards to
resolved peaks. (b) QPT due to an abrupt spin-splitting of the peaks. }
\end{figure}

Fig.\ref{fig:PPic6} illustrates how spin-resolved DOS of the adatoms
depend on the Fano parameter. The variation of $q_{0}$ in the wide
range below the critical value $q_{c1}\approx1.1766$ shifts the position
of the peaks corresponding to opposite spin components equally, as
it is shown in the upper panel. However, above the critical value
the spin splitting abruptly appears as it is shown at the lower panel
of the figure, which clearly indicates that the system undergoes a
QPT. The abrupt appearance of the spin splitting is intimately connected
with the step-like behavior observed in the occupation numbers shown
in Fig.\,\ref{fig:PPic2}. Note that the increase of the Fano factor
above the second critical value $q_{c2}$ leads to the recovering
of the spin-degeneracy as the regime of the phantom atom with cubic
scaling of the pseudogap $\Delta\propto|\varepsilon|^{3}$ is achieved.

Within the critical boundaries $q_{c1}<q_{0}<q_{c2},$ the quantity
$\text{{\tt Re}}({{\tilde{\Sigma}}^{\bar{\sigma}}}_{j\bar{j}}-\Sigma)\equiv\mathcal{J}^{\text{{exch}}}$
from Eq.\,(\ref{eq:TSE}) plays the role of a Zeeman-like splitting
of the levels $\varepsilon_{d}$ in the adatoms. This splitting arises
from an intrinsic exchange field $\mathcal{J}^{\text{{exch}}}$ between
the adatoms intermediated by the graphene monolayer. Its value is
ruled by the system natural control parameter, namely the Fano factor
$q_{0},$ which drives the graphene system towards a QPT. As the upper
and lower adatoms magnetize equally, cf. Fig.\,\ref{fig:PPic2},
the coupling between them is revealed as ferromagnetic. Note that
the dependence of the effective field on the Fano parameter is non-monotonous:
it drops abruptly when $q_{0}=q_{c2}\approx1.3582.$

\section{Conclusions}

In summary, we have proposed a setup based on graphene-adatom system
in which magnetic BICs are triggered by a quantum phase transition
in the region of the quadratic scaling of the pseudogap with energy,
$\Delta\propto|\varepsilon|^{2}$. The control parameter which drives
this transition is a Fano factor of interference tunable by changing
the slope of the Dirac cones in graphene band-structure.

\section{Acknowledgments}

This work was supported by CNPq, CAPES, 2014/14143-0 S{ã}o Paulo
Research Foundation (FAPESP), RISE project 644076 CoExAN, FP7 IRSES project QOCaN and Rannis project
``Bose and Fermi systems for spintronics''. A.\,C.\,S. thanks the NTU at Singapore for hospitality.

\appendix*

\section{LDOS derivation}

To obtain the analytical expressions of the LDOS given by Eq.(\ref{eq:FM_LDOS})
appearing in the conductance of Eq.(\ref{eq:Conductance}), we begin
by applying the equation-of-motion approach to $\mathcal{G}_{\sigma}=-\frac{i}{\hbar}\theta\left(\tau\right){\tt Tr}\{\varrho_{\text{2D}}[\Psi_{\sigma}(\tau),\Psi_{\sigma}^{\dagger}(0)]_{+}\},$
with Eq.(\ref{eq:FOperator}) rewritten as

\begin{align}
\Psi_{\sigma} & =\frac{1}{2\pi}\sqrt{\frac{\pi\Omega_{0}}{\mathcal{N}}}\sum_{s}\int\sqrt{\left|k\right|}(1-q_{0}\frac{\hbar v_{F}k}{D})dkc_{sk\sigma}\nonumber \\
 & +(\pi\text{\ensuremath{\mathcal{D}}}_{0}v_{0})\sum_{j}\mathcal{C}_{j}d_{j\sigma},\label{eq:FOperator2}
\end{align}
expressed in terms of $\mathcal{C}_{j}=(\pi\text{\ensuremath{\mathcal{D}}}_{0}v_{0})^{-1}(t_{d_{1}}/t_{c})\delta_{j1}.$
Substituting Eq. (\ref{eq:FOperator2}) in $\mathcal{G}_{\sigma}$,
one finds
\begin{align}
\mathcal{G}_{\sigma} & =\left(\frac{1}{2\pi}\sqrt{\frac{\pi\Omega_{0}}{\mathcal{N}}}\right)^{2}\sum_{s\tilde{s}}\int\sqrt{\left|k\right|}(1-q_{0}\frac{\hbar v_{F}k}{D})dk\nonumber \\
 & \times\sqrt{\left|q\right|}(1-q_{0}\frac{\hbar v_{F}q}{D})dq\mathcal{G}_{c_{sk\sigma}c_{\tilde{s}q\sigma}}+(\pi\text{\ensuremath{\mathcal{D}}}_{0}v_{0})\sum_{js}\mathcal{C}_{j}\nonumber \\
 & \times\left(\frac{1}{2\pi}\sqrt{\frac{\pi\Omega_{0}}{\mathcal{N}}}\right)\int\sqrt{\left|k\right|}(1-q_{0}\frac{\hbar v_{F}k}{D})dk\nonumber \\
 & \times(\mathcal{G}{}_{d_{j\sigma}c_{sk\sigma}}+\mathcal{G}_{c_{sk\sigma}d_{j\sigma}})+(\pi\text{\ensuremath{\mathcal{D}}}_{0}v_{0})^{2}\sum_{jl}\mathcal{C}_{j}\mathcal{C}_{l}\mathcal{G}_{d_{j\sigma}d_{l\sigma},}\nonumber \\
\label{eq:GF_1}
\end{align}
with the new Green's functions $\mathcal{G}_{c_{sk\sigma}c_{\tilde{s}q\sigma}}$,
$\mathcal{G}_{d_{j\sigma}c_{sk\sigma}}$, $\mathcal{G}_{c_{sk\sigma}d_{j\sigma}}$
and $\mathcal{G}_{d_{j\sigma}d_{l\sigma}}$ to be determined. To this
end, we first consider $\mathcal{G}_{c_{sk\sigma}c_{\tilde{s}q\sigma}}=-\frac{i}{\hbar}\theta\left(\tau\right){\tt Tr}\{\varrho_{\text{2D}}[c_{sk\sigma}\left(\tau\right),c_{\tilde{s}q\sigma}^{\dagger}\left(0\right)]_{+}\},$
whose time derivative $\partial_{\tau}\equiv\frac{\partial}{\partial\tau}$
gives
\begin{eqnarray}
\partial_{\tau}\mathcal{G}_{c_{sk\sigma}c_{\tilde{s}q\sigma}} & = & -\frac{i}{\hbar}\delta\left(\tau\right){\tt Tr}\{\varrho_{\text{2D}}[c_{sk\sigma}\left(\tau\right),c_{\tilde{s}q\sigma}^{\dagger}\left(0\right)]_{+}\}\nonumber \\
 & - & \frac{i}{\hbar}(\hbar v_{F}k)\mathcal{G}_{c_{sk\sigma}c_{\tilde{s}q\sigma}}-\frac{i}{\hbar}\sum_{j}\mathcal{V}_{k}\mathcal{G}_{d_{j\sigma}c_{\tilde{s}q\sigma}},\nonumber \\
\label{eq:GF_3}
\end{eqnarray}
where we have used
\begin{align}
i\hbar\partial_{\tau}c_{sk\sigma}\left(\tau\right) & =[c_{sk\sigma},\mathcal{H}_{\text{2D}}]=(\hbar v_{F}k)c_{sk\sigma}\left(\tau\right)\nonumber \\
 & +\sum_{j}\mathcal{V}_{k}d_{j\sigma}\left(\tau\right).\label{eq:HB_I}
\end{align}

In the energy domain after performing the time Fourier transform,
we solve Eq. (\ref{eq:GF_3}) for $\tilde{\mathcal{G}}_{c_{sk\sigma}c_{\tilde{s}q\sigma}}$
and obtain

\begin{align}
\tilde{\mathcal{G}}_{c_{sk\sigma}c_{\tilde{s}q\sigma}} & =\frac{\delta\left(k-q\right)\delta_{s\tilde{s}}}{\varepsilon^{+}-\hbar v_{F}k}+\sum_{j}\frac{\mathcal{V}_{k}}{\varepsilon^{+}-\hbar v_{F}k}\tilde{\mathcal{G}}_{d_{j\sigma}c_{sq\sigma}}.\label{eq:GF_4}
\end{align}

Notice that we also need to calculate the mixed Green's function $\tilde{\mathcal{G}}_{d_{j\sigma}c_{sq\sigma}}.$
We then define the advanced Green's function $\mathcal{F}_{d_{j\sigma}c_{sq\sigma}}=\frac{i}{\hbar}\theta\left(-\tau\right){\tt Tr}\{\varrho_{\text{2D}}[d_{j\sigma}^{\dagger}\left(0\right),c_{sq\sigma}\left(\tau\right)]_{+}\},$
whose equation-of-motion reads,
\begin{align}
\partial_{\tau}\mathcal{F}_{d_{j\sigma}c_{sq\sigma}} & =-\frac{i}{\hbar}\delta\left(\tau\right){\tt Tr}\{\varrho_{\text{2D}}[d_{j\sigma}^{\dagger}\left(0\right),c_{sq\sigma}\left(\tau\right)]_{+}\}\nonumber \\
 & -\frac{i}{\hbar}(\hbar v_{F}q)\mathcal{F}_{d_{j\sigma}c_{sq\sigma}}-\frac{i}{\hbar}\sum_{l}\mathcal{V}_{q}\mathcal{F}_{d_{j\sigma}d_{l\sigma}},\label{eq:GF_6}
\end{align}
where we have used once again Eq. (\ref{eq:HB_I}), interchanging
$k\leftrightarrow q$. The Fourier transform of Eq.~(\ref{eq:GF_6})
leads to
\begin{align}
\varepsilon^{-}\tilde{\mathcal{F}}_{d_{j\sigma}c_{sq\sigma}} & =(\hbar v_{F}q)\tilde{\mathcal{F}}_{d_{j\sigma}c_{sq\sigma}}+\sum_{l}\mathcal{V}_{q}\tilde{\mathcal{F}}_{d_{j\sigma}d_{l\sigma}},\label{eq:GF_7}
\end{align}
with $\varepsilon^{-}=\varepsilon-i0^{+}$. Applying the property
$\tilde{\mathcal{G}}_{d_{j\sigma}c_{sq\sigma}}=(\tilde{\mathcal{F}}_{d_{j\sigma}c_{sq\sigma}})^{\dagger}$
on Eq. (\ref{eq:GF_7}), we show that
\begin{align}
\varepsilon^{+}\tilde{\mathcal{G}}_{d_{j\sigma}c_{sq\sigma}} & =(\hbar v_{F}q)\tilde{\mathcal{G}}_{d_{j\sigma}c_{sq\sigma}}+\sum_{l}\mathcal{V}_{q}\tilde{\mathcal{G}}_{d_{j\sigma}d_{l\sigma}},\label{eq:GF_8}
\end{align}

\begin{equation}
\tilde{\mathcal{G}}_{d_{j\sigma}c_{sq\sigma}}=\sum_{l}\frac{\mathcal{V}_{q}}{\varepsilon^{+}-\hbar v_{F}q}\tilde{\mathcal{G}}_{d_{j\sigma}d_{l\sigma}}\label{eq:GF_9}
\end{equation}
and analogously,
\begin{equation}
\tilde{\mathcal{G}}_{c_{sq\sigma}d_{j\sigma}}=\sum_{l}\frac{\mathcal{V}_{q}}{\varepsilon^{+}-\hbar v_{F}q}\tilde{\mathcal{G}}_{d_{l\sigma}d_{j\sigma}}.\label{eq:GF_10}
\end{equation}
Now we substitute Eq. (\ref{eq:GF_9}) into Eq. (\ref{eq:GF_4}) and
the latter, together with Eqs. (\ref{eq:GF_10}) and (\ref{eq:SE})
for the self-energy splitted as

\begin{equation}
\Sigma=\sum_{s}\int dk\frac{\mathcal{V}_{k}\mathcal{V}_{k}}{\varepsilon^{+}-\hbar v_{F}k}=\pi v_{0}^{2}\text{\ensuremath{\mathcal{D}}}_{0}(\tilde{\mathcal{A}}_{j}-i\mathcal{B}_{j}),\label{eq:SEsplitted}
\end{equation}
into Eq. (\ref{eq:GF_1}) in the energy domain, which results in

\begin{align}
\tilde{\mathcal{G}}_{\sigma} & =\left(\frac{1}{2\pi}\sqrt{\frac{\pi\Omega_{0}}{\mathcal{N}}}\right)^{2}\sum_{s}\int(1-q_{0}\frac{\hbar v_{F}k}{D})^{2}kdk\frac{1}{\varepsilon^{+}-\varepsilon_{k}}\nonumber \\
 & +(\pi\text{\ensuremath{\mathcal{D}}}_{0}v_{0})^{2}\sum_{jl}(\mathcal{\tilde{A}}_{j}-i\mathcal{B}_{j})\tilde{\mathcal{G}}_{d_{j\sigma}d_{l\sigma}}(\mathcal{\tilde{A}}_{l}-i\mathcal{B}_{l})\nonumber \\
 & +(\pi\text{\ensuremath{\mathcal{D}}}_{0}v_{0})^{2}\sum_{jl}\mathcal{C}_{j}(\mathcal{\tilde{A}}_{l}-i\mathcal{B}_{l})(\tilde{\mathcal{G}}_{d_{j\sigma}d_{l\sigma}}+\tilde{\mathcal{G}}_{d_{l\sigma}d_{j\sigma}})\nonumber \\
 & +(\pi\text{\ensuremath{\mathcal{D}}}_{0}v_{0})^{2}\sum_{jl}\mathcal{C}_{j}\mathcal{C}_{l}\tilde{\mathcal{G}}_{d_{j\sigma}d_{l\sigma}}.\label{eq:GFN}
\end{align}
Thus after some algebra via the evaluation of $-\frac{1}{\pi}\sum_{\sigma}{\tt Im}(\tilde{\mathcal{G}}_{\sigma})$,
we determine Eq.(\ref{eq:FM_LDOS}) as the LDOS probed by the STM
tip, with

\begin{align}
\Delta\text{{LDOS}}{}_{jl\sigma} & =-(\pi v_{0}^{2}\text{\ensuremath{\mathcal{D}}}_{0}^{2}){\tt Im}[(\mathcal{A}{}_{l}-i\mathcal{B}_{l})\tilde{\mathcal{G}}_{d_{l\sigma}d_{j\sigma}}(\mathcal{A}{}_{j}-i\mathcal{B}_{j})],\nonumber \\
\label{eq:LDOSp1}
\end{align}
with $\mathcal{A}{}_{j}=\frac{1}{\pi v_{0}^{2}\text{\ensuremath{\mathcal{D}}}_{0}}\text{{\tt Re}}\Sigma+\delta_{j1}(\pi^{2}v_{0}^{2}\text{\ensuremath{\mathcal{D}}}_{0}^{2})^{-1/2}(t_{d1}/t_{c})$
and $\mathcal{B}_{j}=-\frac{1}{\pi v_{0}^{2}\text{\ensuremath{\mathcal{D}}}_{0}}\text{{\tt Im}}\Sigma.$

\end{document}